%% file: ms.tex
\pdfoutput=1

\documentclass[twocolumn]{emulateapj}
\usepackage{graphicx}

\usepackage[z,alpha]{tmsfnts}

\usepackage{color}

\slugcomment{submitted to the Astrophysical Journal}

\shorttitle{Size-virial radius relation of galaxies}
\shortauthors{Kravtsov A.}

\begin{document}

\input{def}

\input{macros}

\setlength{\hbadness}{10000}


\title{The Size-virial radius relation of galaxies}
\author{Andrey V. Kravtsov\altaffilmark{1,2,3}}
  
\altaffiltext{1}{Department of Astronomy \& Astrophysics, The University of Chicago, Chicago, IL 60637 USA; {\tt andrey@oddjob.uchicago.edu}} 
\altaffiltext{2}{Kavli Institute for Cosmological Physics, The University of Chicago, Chicago, IL 60637 USA} 
\altaffiltext{3}{Enrico Fermi Institute, The University of Chicago, Chicago, IL 60637}


\begin{abstract}
Sizes of galaxies are an important diagnostic for galaxy formation models. In this study I use the abundance matching ansatz, which has proven to be successful in reproducing galaxy clustering and other statistics, to derive estimates of the virial radius, $\Rtwoh$, for galaxies of different morphological types and wide range of stellar mass. I show that over eight of orders of magnitude in stellar mass galaxies of all morphological types follow an approximately linear relation between half-mass radius of their stellar distribution, $\rhalf$ and virial radius, $\rhalf\approx 0.015\Rtwoh$ with a scatter of $\approx 0.2$ dex. Such scaling is in remarkable agreement with expectation of models which assume that galaxy sizes are controlled by halo angular momentum, which implies $\rhalf\propto\lambda \Rtwoh$, where $\lambda$ is the spin of galaxy parent halo. The scatter about the relation is comparable with the scatter expected from the distribution of $\lambda$ and normalization of the relation agrees with that predicted by the model of Mo, Mao \& White (1998), if galaxy sizes were set on average at $z\sim 1-2$. Moreover, I show that when stellar and gas surface density profiles of galaxies of different morphological types are rescaled using radius $r_n= 0.015\Rtwoh$, the rescaled surface density profiles follow approximately universal exponential (for late types) and de Vaucouleurs (for early types) profiles with scatter of only $\approx 30-50\%$ at $R\approx 1-3r_n$. Remarkably, both late and early type galaxies have similar mean stellar surface density profiles at $R\gtrsim 1r_n$. The main difference between their stellar distributions is thus at $R<r_n$. The results of this study imply that galaxy sizes and radial distribution of baryons are shaped primarily by properties of their parent halo and that sizes of both late type disks and early type spheroids are controlled by halo angular momentum. 
\end{abstract}

\keywords{}

\section{Introduction}
\label{sec:intro}

In the standard hierarchical structure formation scenario galaxies form at the minima of potential wells formed by nonlinear collapse of peaks in the initial density field. The process of galaxy formation is expected to be complex,
highly nonlinear process, involving forces on a wide range of scales, supersonic, highly compressible and turbulent flows of gas, and a variety of cooling, heating, and feedback processes.  

Despite the apparent complexity of formation processes, observed galaxies exhibit a number of tight scaling relations between their structural parameters and, as first shown by \citet{fall_efstathiou80} and elaborated by \citet*[][see also \citet{dalcanton_etal97,avila_reese_etal98,avila_reese_etal08,dutton_etal07,dutton09,fu_etal10}]{mo_etal98}, these scaling relations can be reproduced in a fairly simple framework, in which sizes of galaxies are determined by the sizes of their initial rotationally supported gaseous disks, which, in turn are set by the angular momentum of gas. Under assumption that gas angular momentum is proportional to that of dark matter, such models predict that galaxy size should scale as $\propto \lambda\Rtwoh$, where $\Rtwoh$ is the ``virial'' radius defined as the radius enclosing overdensity of 200 with respect to the critical density of the universe, $\rho_{\rm cr}(z)$,  so that $M_{200}=(4\pi/3)200\rho_{\rm cr}(z)R^3_{200}$.   

In addition, regularity in galaxy properties is implied by  success of the abundance matching ansatz, in which relation between total mass of halos, $M$, and stellar mass of galaxies they host, $M_*$ is derived from a simple assumption that the relation is approximately monotonic and cumulative abundance of galaxies with masses above a given $M_*$ is matched to cumulative number density of halos with masses above $M$. This model is remarkably successful in reproducing clustering of galaxies of different luminosities and at different redshifts  \citep{kravtsov_etal04,tasitsiomi_etal04,conroy_etal06,reddick_etal12}
and other statistics  \citep{vale_ostriker04,vale_ostriker06,behroozi_etal10,behroozi_etal12,guo_etal10,moster_etal12,hearin_etal12}.

In this paper, I use the abundance matching ansatz to examine relation between sizes of stellar systems of galaxies, characterized by the {\it three-dimensional} half-mass radius, $\rhalf$ and virial radius of their halos, $\Rtwoh$, derived using the abundance matching ansatz. I show that over the entire observed range of stellar masses and morphologies, galaxies exhibit an approximately linear scaling relation between stellar half-mass radius and halo virial radius with normalization and scatter consistent with expectation of the \citet{mo_etal98} model. Furthermore, I show that stellar and gas surface density profiles of galaxies  rescaled using radius $r_n=0.015\Rtwoh$ follow universal  profiles with a scatter as low as $\approx 30-50\%$ at intermediate radii within optical extent of galaxies. 

Throughout this paper I assume a flat $\Lambda$CDM model with parameters $\Omega_\rmm=1-\Omega_\Lambda=0.27$, $\Omega_\rmb=0.0469$, $h=H_0/(100\kmsmpc)=0.7$, $\sigma_8=0.82$ and $n_\rms=0.95$ compatible with combined constraints from WMAP, BAO, SNe, and cluster abundance \citep{komatsu_etal11}.

\section{Abundance matching}
\label{sec:am}

To estimate the virial masses and radii of halos hosting galaxies, I use the abundance matching ansatz, in which relation between total halo mass, $M$, and stellar mass of galaxies they host, $M_*$, is established implicitly by matching cumulative stellar and halo mass functions: $n_{\rm h}(>M)=n_{\rm g}(>M_*)$. 

A number of estimates of the $M_*-M$ relation using this technique has been presented in the recent literature \citep[e.g.,][]{moster_etal12,behroozi_etal12}. However, the relations derived in these studies are based on stellar mass functions (SMFs) known to underestimate abundance of massive galaxies \citep{bernardi_etal10} and the double power law fit to the $M_*-M$ relation of \citet{moster_etal12} does not capture the upturn in the relation at $M_*\lesssim 10^9\ M_{\odot}$ originating from the steepening of the stellar mass function at these masses \citep{baldry_etal08,baldry_etal12,papastergis_etal12}. Therefore, in this study I re-derive the $M_*-M$ relation to fix these problems.
 
I use the \citet{tinker_etal08} calibration of the halo mass function for $\Mtwoh$, which was calibrated using host halos only. To account for subhalos, I correct the host mass function by subhalo fraction, $f_{\rm sub}(>M)=[n_{\rm tot}(>M)-n_{\rm host}(>M)]/n_{\rm host}(>M)$, to get $n_{\rm tot}(>M)$ -- the mass function that includes both hosts and subhalos. The latter was calculated using current $\Mtwoh$ masses for hosts and corresponding masses at the accretion epoch for subhalos using $z=0$ halo catalog halo catalog of \citet{behroozi_etal11} derived from the Bolshoi simulation \citep{klypin_etal11} of $(250h^{-1}\rm Mpc)^3$ volume in the concordance cosmology adopted in this study. The subhalo fraction in the Bolshoi simulation is parametrized as $f_{\rm sub}={\rm min}[0.35,0.085(15-\log_{10}\Mtwoh)]$. The halo mass function derived from the Bolshoi simulation agrees within $5\%$ with the \citet{tinker_etal08} parameterization, but the latter is more accurate at the highest halo masses. 

I combine two recent calibrations of the SMF by \citet{papastergis_etal12} and \citet{bernardi_etal10} to accurately  characterize SMF behavior at both small and large $M_*$, respectively. I use these two calibrations to construct a combined stellar mass function, $n(M_*)={\rm max}[n_{\rm P12},n_{\rm B10}]$, that spans from $M_*\approx 10^7\rm \Msun$ to $M*\approx 10^{12}\ \Msun$. Both stellar mass functions assume \citet{chabrier03} IMF to estimate stellar masses of galaxies. For $n_{\rm P12}$ I adopt double Schechter form given by eq.~6 of \citet{baldry_etal12} with the following parameters: $\log_{10}M_*=10.66$, $\phi_1^{\ast
}=3.96\times 10^{-3}$, $\alpha_1=-0.35$, $\phi_2^{\ast}=6.9\times 10^{-4}$, $\alpha_2=-1.57$. These parameters are in general agreement with the best fit parameters derived for the local stellar mass function by \citet{baldry_etal12}. Note that SMF at $M_*\lesssim 10^8\ \Msun$ is quite uncertain due to incompleteness of low surface brightness galaxies in this regime \citep{baldry_etal12}; the current SMF measurements at these stellar masses should be considered as lower limits and the actual SMF may be somewhat steeper still.    For $n_{\rm B10}$ I use  parameter values given in the bottom row of Table 4 in \citet[][unbracketed values]{bernardi_etal10} and the Schechter parametrization of the SMF given by eq.~9 in that paper. I refer readers to the original papers for further details on how the stellar mass functions were estimated.

\section{Galaxy samples}
\label{sec:samples}

To estimate the size--virial radius relation, I have selected several publicly available datasets chosen to span the entire range of galaxy stellar masses\footnote{Stellar masses in all of the samples were estimated assuming the \citet{chabrier03} IMF.} and morphologies. First, I use a compilation of stellar masses and effective radii for spheroidal, early-type galaxies from \citet{misgeld_hilker11}. These include ellipticals (Es) and dwarf elliptical (dE) galaxies in the Virgo cluster with HST \citep{ferrarese_etal06} and the VLT/FORS1 observations of dEs in the Hydra I and Centaurus clusters \citep{misgeld_etal08,misgeld_etal09}, and the dwarf spheroidal (dSph) galaxies in the Local Group. The sample of late type galaxies includes the THINGS/HERACLES galaxies of \citet{leroy_etal08} and the LITTLE THINGS sample of dwarf irregular galaxies from \citet{zhang_etal12}. I also include the stellar mass profile of the Milky Way using a combination of thin and thick stellar disks with parameters given in Table 2 of \citet{mcmillan11}. For the late type samples, I used the deprojected stellar surface density profiles presented in these studies to estimate the half mass radius, $\rhalf$, directly from profiles. The radius $\rhalf$ was determined as the radius that contains half of the stellar mass of galaxies using the cumulative mass profile of each disk: $M_*(<R)=2\pi\int_0^R\Sigma(R^\prime)R^\prime dR^\prime$.

In addition, I use the average relations between half-light radius and stellar mass, $\langle \rhalf\vert M_*\rangle$, derived for early and late type galaxies in the SDSS  from the recent study by \citet[][SerExp values in their Table 4]{bernardi_etal12}. 
I also use intrinsic scatter about the mean relation calculated for both early and late type galaxies (M. Bernardi 2012, priv. communication). Finally, I use a half-mass radii and stellar masses for a sample of massive SDSS galaxies presented in \citet[][see their Table 1]{szomoru_etal12}.

\begin{figure}[t!]
{\centering
\includegraphics[trim=15 0 0 0,scale=0.475]{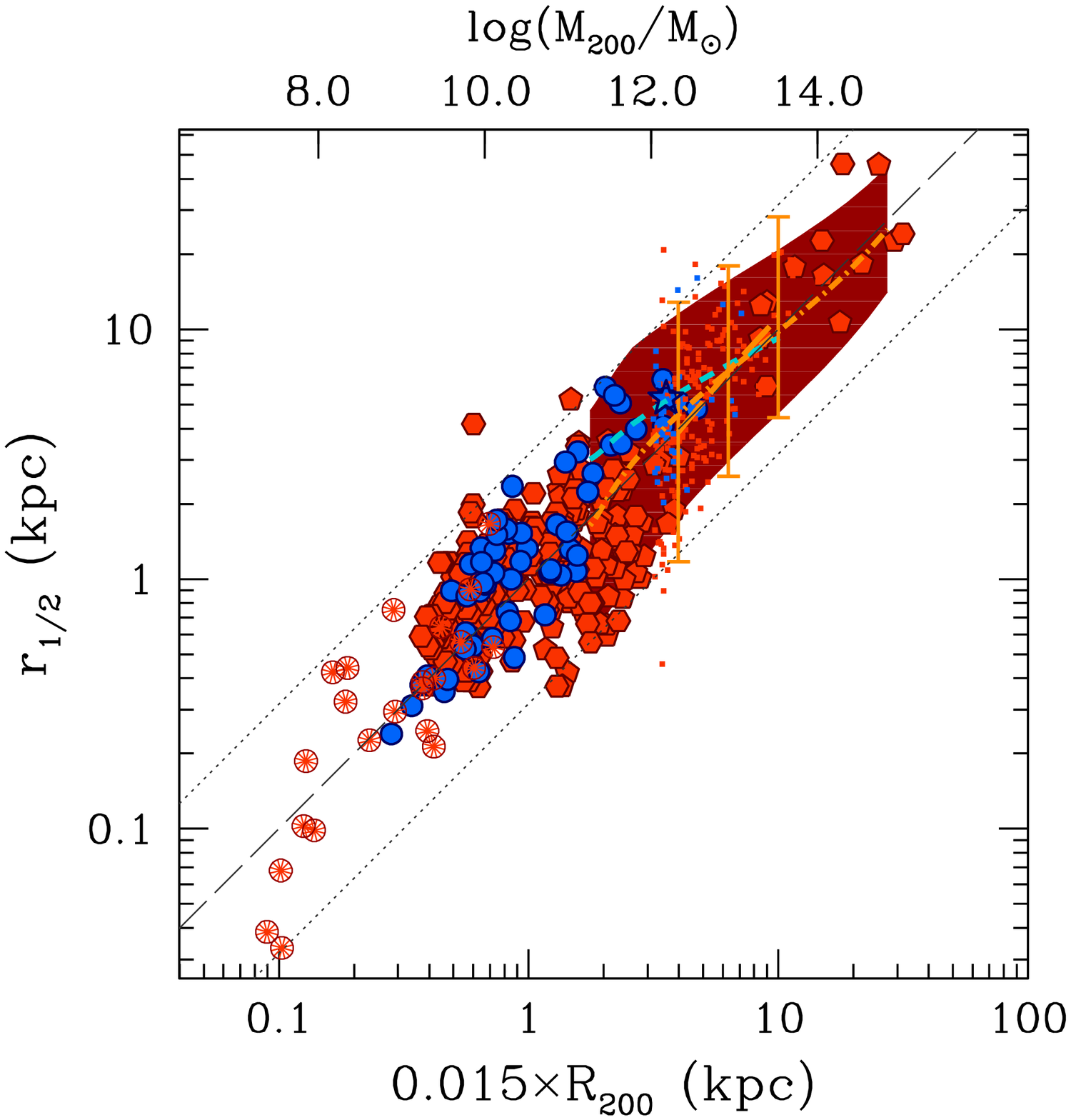}
}
\caption{Relation between half-mass radius of stellar distribution in galaxies of different stellar masses (spanning more than eight orders of magnitude in stellar mass) and morphological types and inferred virial radius of their parent halos, $\Rtwoh$, defined as the radius enclosing overdensity of $200\rhoc$, and estimated as described in \S~\ref{sec:am}. The {\it red pentagons} and {\it hexagons} show a sample of elliptical and dwarf elliptical galaxies from the compilation of \protect\citet{misgeld_hilker11}; {\it blue circles} are the late type galaxies from the samples of \protect\citet{leroy_etal08} and \citet{zhang_etal12} with half-mass radii estimated as described in \S~\ref{sec:samples}, while the {\it star symbol} shows the Milky Way; the {\it red cartwheel} points show the Local Group dwarf spheroidal galaxies from the compilation of \protect\citet{misgeld_hilker11}. The {\it light blue dashed line} and {\it dot-dashed orange line} show the average relations derived for late and early-type galaxies, respectively, from the average $R_{\rm 1/2}-M_*$ relations of \protect\citet{bernardi_etal12}. {\it Dark red shaded band} shows $2\sigma$ scatter around the mean relation calculated for all galaxies in the \citet{bernardi_etal12} sample. {\it The orange dot-dashed line} with error bars shows the mean relation and $2\sigma$ scatter for massive SDSS galaxies presented in \protect\citet{szomoru_etal12}; individual galaxies from this sample are shown by blue (S\'ersic index $n<2.5$) and red ($n>2.5$) dots. {\it The gray dashed line} shows linear relation $r_{1/2}=0.015\Rtwoh$ and dotted lines are linear relations offset by 0.5 dex, which approximately corresponds to the scatter in galaxy sizes from distribution of  halo spin parameter $\lambda$ under assumption that $r_{1/2}\propto\lambda\Rtwoh$.}
\label{fig:rhR200}
\end{figure}

\section{Results} 
\label{sec:results}

\subsection{The size-virial radius relation}
\label{subsec:rhrn}

To derive the size-virial radius relation, I first assign $M_{200}$ to galaxies using their stellar mass and the $M_*-M_{200}$ relation derived using abundance matching. I then estimate the {\it three-dimensional} half-mass radius from the projected 2D half-mass radii reported for observed galaxies. 

I assume that in late type galaxies stars are in a disk and hence the 2D $R_{1/2}$ radius is equal to the 3D $\rhalf$ radius. For early  type galaxies I assume that stars have spheroidal distribution and   convert projected $\Reff$ into the 3D half-mass radius using $\rhalf=1.34\Reff$. This expression is accurate for spheroidal systems described by the S\'ersic profile with a wide range of the S\'ersic index values \citep[see eq. 21 in][]{limaneto_etal99}. 

\begin{figure}[t!]
{\centering
\includegraphics[trim=15 0 0 60,scale=0.475]{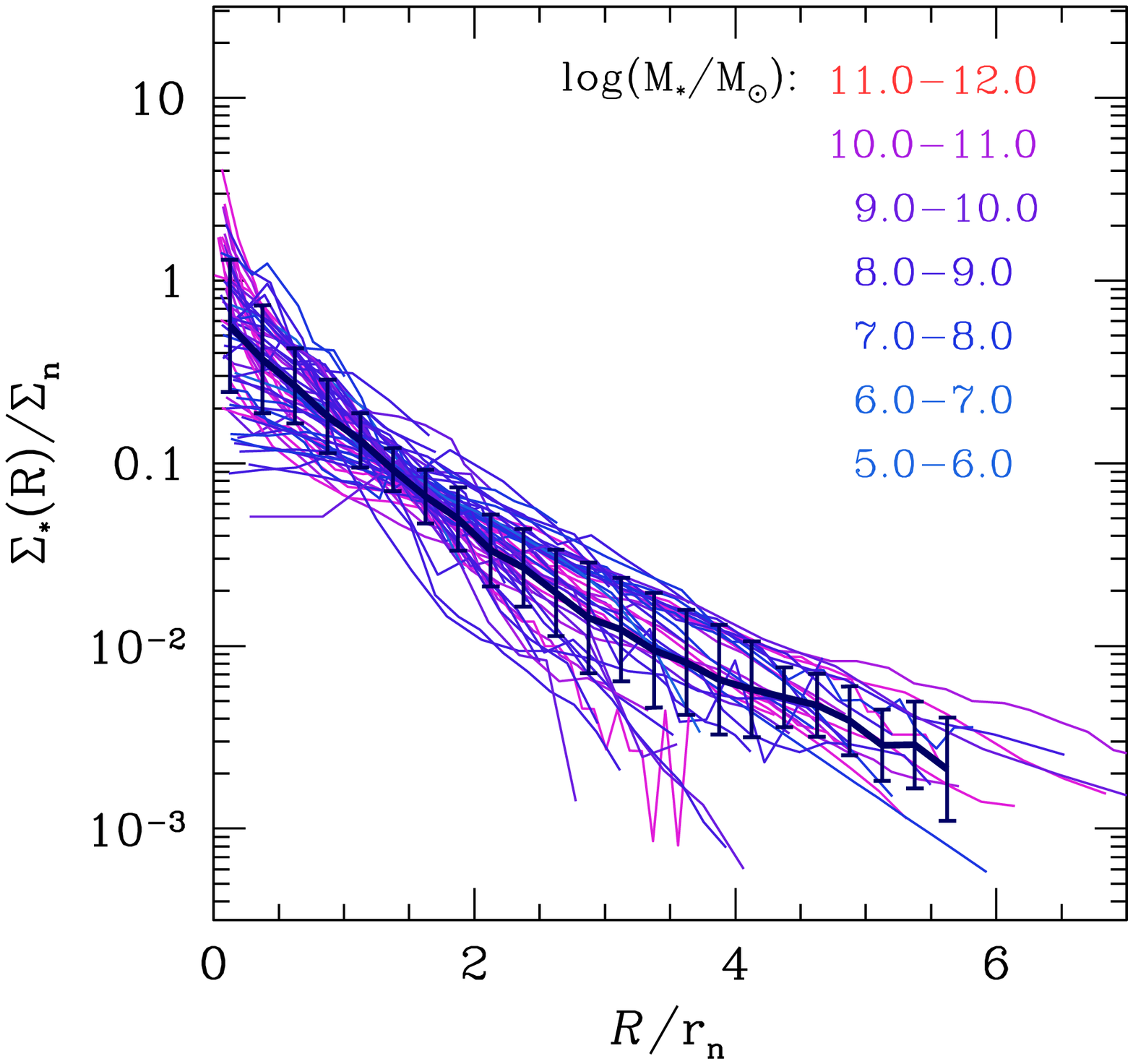}
\includegraphics[trim=15 0 0 85,scale=0.475]{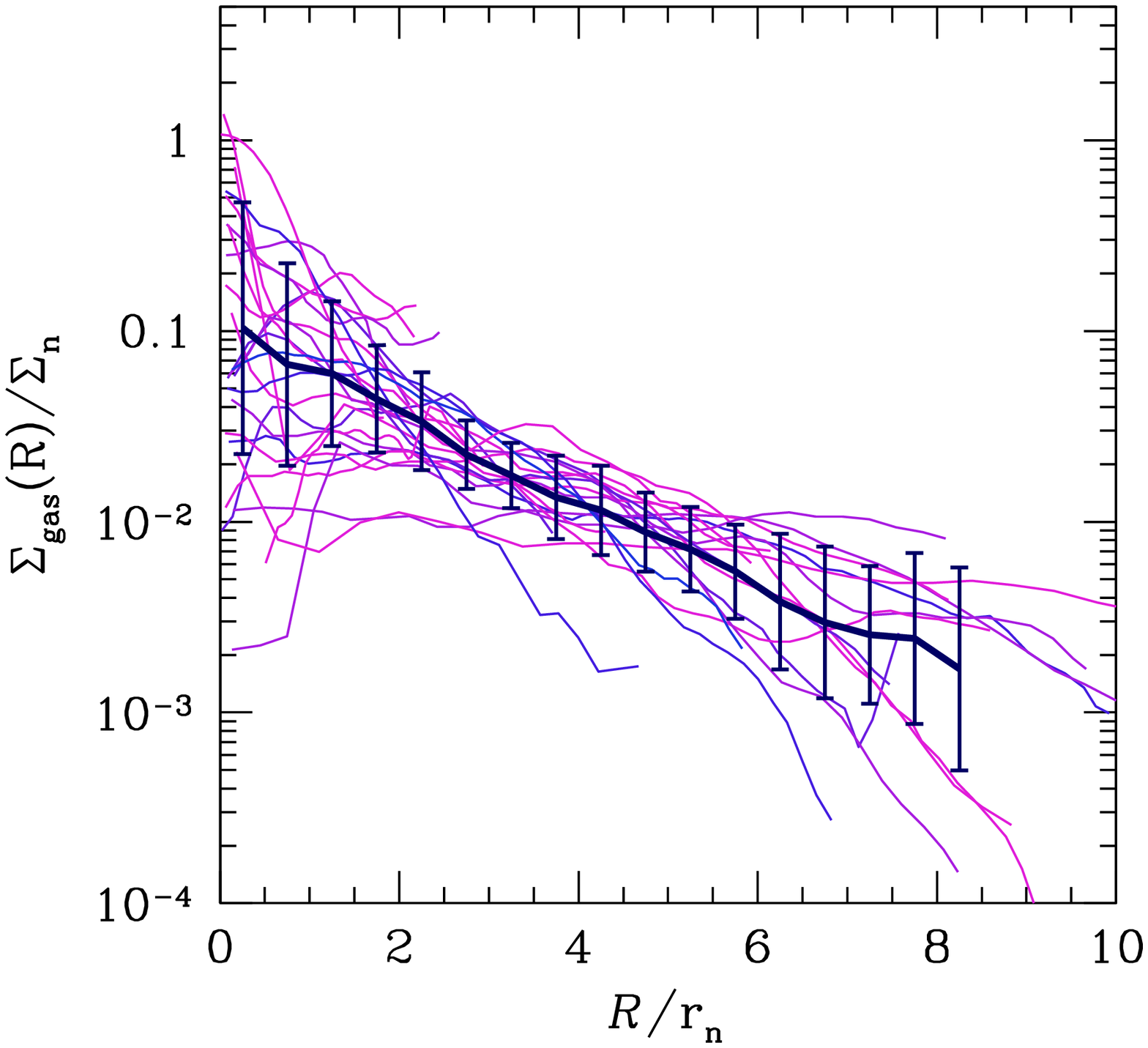}
\vspace{-1cm}
}
\caption{Normalized surface density profiles of stars, $\Sigma_*(R)/\Sigma_{n,*}$ (top panel) and neutral gas, $\Sigma_{\rm gas}(R)/\Sigma_{n,g}$ (HI$+$H$_2$, bottom panel) for late type galaxies. The top panel includes galaxies from the \protect\citet{leroy_etal08} and the LITTLE THINGS sample of \protect\citet{zhang_etal12}, while the bottom panel only includes the surface density profiles of gas from \protect\citet{leroy_etal08}. Profiles of individual galaxies are shown by thin lines colored according to their $\log_{10}M_*$ as indicated in the legend. Each individual profile is normalized by the radius $r_n=0.015R_{200}$, where $R_{200}$ is obtained using the abundance matching ansatz from the galaxy $M_*$. The thick lines with errorbars show the sample average and rms dispersion around the mean. Both stellar gas profiles are well described by the exponential profile on average, but the scale length of the gas profile is on average a factor of $\approx 2.7$ larger than that of the stellar profile.}
\label{fig:sspro}
\end{figure}
Figure~\ref{fig:rhR200} shows the derived relation between $\rhalf$ and the virial radius of parent halos, $\Rtwoh$, for all the observational samples described above. Remarkably, $\rhalf$ scales approximately linearly  over two orders of magnitudes in radius and over eight orders of magnitude in stellar mass from the dwarf spheroidal galaxies to the most massive ellipticals, as expected in the models assuming that size of galaxies is proportional to $\lambda\Rtwoh$ \citep{fall_efstathiou80,mo_etal98}. The linear relation, $\rhalf=0.015\Rtwoh$ is shown by the gray dashed line. The normalization of this relation is chosen so that distribution of points is approximately symmetric around the line. The formal power law fit to the $\rhalf-\Rtwoh$ relation of individual galaxy points shown in the figure gives slope of $0.95\pm 0.065$, normalization of $0.015\pm 0.0007$, and scatter of $0.20\pm 0.016$ dex (all errors indicating 95\% confidence interval). These values should be taken only as indicative that relation indeed has the slope close to linear, given that galaxy sample only samples the mass range in a semi-random fashion. More robust determination of the slope, normalization, and scatter will require galaxy samples with uniform, well-defined selection criteria. 

As I discuss in \S~\ref{sec:conclusions} below, the normalization of the $\rhalf-\Rtwoh$ relation is {\it quantitatively} consistent with the expectations of the \citet{mo_etal98} model if we assume that angular momentum of the disk is set near the end of fast mass accretion epoch at $z\sim 2$ and take into account pseudo-evolution of halo mass \citep{diemer_etal12}.

It is noteworthy that dIrr galaxies have similar 
half-mass radii to the dwarf ellipticals and follow approximately the same $\rhalf-\Rtwoh$ relation. Massive late type galaxies also follow the same linear relation, although the figure indicates that late type galaxies of intermediate stellar mass have systematically larger half-mass radii than early type galaxies of the same stellar mass \citep[e.g.,][]{bernardi_etal12}. 

 The shaded band around dot-dashed line in Fig.~\ref{fig:rhR200} shows $2\sigma\approx 0.3-0.5$ dex {\it intrinsic} scatter estimated for all galaxies in the sample of \citet[][the scatter shown is for all galaxies in the sample, M. Bernardi, priv. comm.]{bernardi_etal12}. The orange error bars show scatter estimated for the mass limited sample of massive SDSS galaxies presented in \citet{szomoru_etal12}. The scatter estimated for this sample is in good agreement with the scatter of the \citet{bernardi_etal12} sample. Remarkably, the scatter is also approximately consistent with the scatter expected from the distribution of halo spins, $\lambda$, in models in which galaxy size is $\propto\lambda\Rtwoh$, shown by the dotted lines in the figure. 
 
\subsection{Stellar and gas surface density profiles of galaxies}
\label{subsec:spro}

 In this section I show that in addition to $r_{1/2}-\Rtwoh$ correlation the surface density profiles of stars and neutral gas approximately follow universal profiles when scaled by $r_n=0.015\Rtwoh$, i.e. the mean normalization of the $\rhalf-\Rtwoh$ correlation.  

Two panels in Figure~\ref{fig:sspro} show the surface density profiles of stars and neutral gas (HI$+$H$_2$, where HI is corrected for helium) for late type galaxies. The radius of each individual profile was rescaled by $r_n=0.015\Rtwoh$ and surface densities were scaled by $\Sigma_n=0.448M/r_n^2$, where $M$ is total stellar or gas mass of each galaxy and factor $0.448=1.678^2/(2\pi)$ assumes exponential profile ($\rhalf=1.678R_d$). The figure shows that both the mean stellar and gas profiles are on average well described by the exponential profile, $\Sigma(R)=\Sigma_0\exp(-R/R_d)$, where for stars $\Sigma_{0,*}\approx 1256M_*/R^2_{200}$ and the scale length $R_{\rm d,*}\approx 0.011\Rtwoh$ and for gas $\Sigma_{0,\rm gas}\approx 199M_{\rm gas}/R^2_{200}$ and $R_{\rm d,gas}\approx 0.029\Rtwoh$.  The gas distribution is thus on average a factor of $\approx 2.6$ more extended than stellar distribution.  
 The scatter around the mean profiles is rather small and is only $\approx 30-50\%$ at $R\sim 1-3r_n$, even though galaxies shown in the $\Sigma_*(R)$ figure (top panel) span over six orders of magnitude in $M_*$. 
 
 The approximate universality of the gas surface density profiles was recently pointed out by \citet{bigiel_blitz12}. These authors rescaled gas profiles of the THINGS/HERACLES galaxies using the optical Holmberg radius, $R_{25}$, and surface density  $\Sigma_{\rm trans}$ at the radius where $\Sigma_{\rm H_2}=\Sigma_{\rm HI}$. Such rescaling results in average exponential profile described by $\Sigma_{\rm gas}=2.1\Sigma_{\rm trans}\exp(-1.65R/R_{25})$ with comparable scatter around the mean profile to  the rescaling described above. Comparison with the scaling derived above gives $\Sigma_{\rm trans}=95M_{\rm gas}/R^2_{200}$ and $R_{25}=0.048\Rtwoh$. Thus, results of \citet{bigiel_blitz12} can be understood if $\Sigma_{\rm trans}$ scales with characteristic surface density $\Sigma_{\rm 0, gas}$. The scaling of $R_{25}$ is implied by the scaling $R_{\rm d,*}\propto R_{200}$ because for exponential disks $R_{25}\approx 4.5R_{\rm d,*}$. Thus, the gas surface density profiles can be scaled by surface density $\propto M_{\rm gas}/R^2_{25}$ instead of $\Sigma_{\rm trans}$. In summary, results presented here indicate that the reason scaling employed by \citet{bigiel_blitz12} works is that surface densities of gas and stars are both exponential and their scale lengths are correlated: $R_{\rm d, gas}\approx R_{\rm d,*}$. The origin of the universality of the $\Sigma_{\rm gas}$ profiles lies in the scaling of half-mass radius of both gas and stars with the virial radius of parent halo. 

\begin{figure}[t!]
{\centering
\includegraphics[trim=15 0 0 60,scale=0.475]{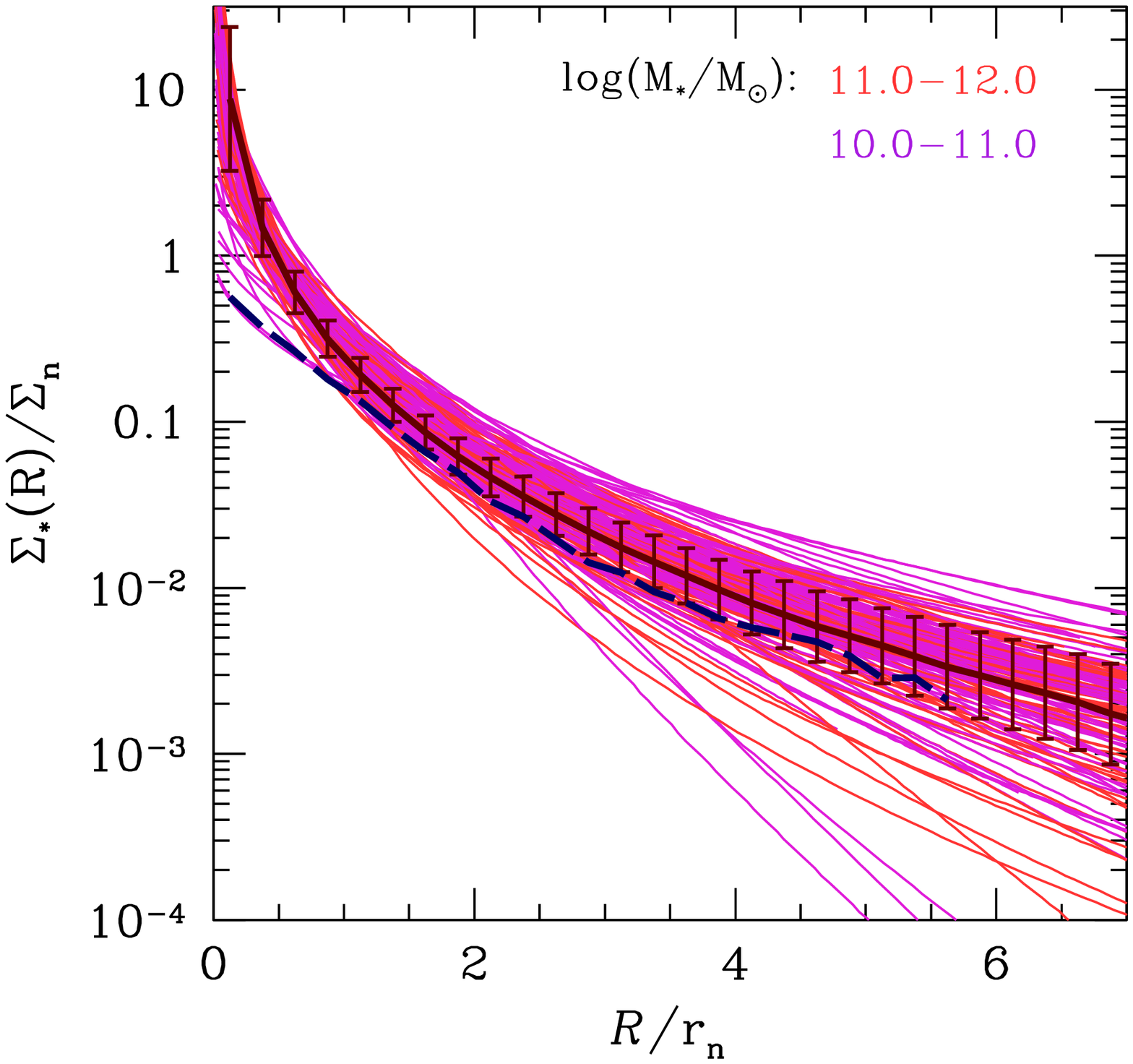}
\vspace{-0.5cm}
}
\caption{Normalized surface density profiles of stars for massive early type SDSS galaxies from \protect\citet{szomoru_etal12}.  Profiles of individual galaxies are shown by thin lines colored according to their $\log_{10}M_*$, as indicated in the legend. Each individual profile is normalized by the radius $r_n=0.015R_{200}$, where $R_{200}$ is obtained using the abundance matching ansatz from the galaxy $M_*$. The thick lines with errorbars show the sample average and rms dispersion around the mean. The thick dashed line shows the average profile of late type galaxies from the top panel of Fig.~\ref{fig:sspro} for comparison. }
\label{fig:ssproser}
\end{figure}

Figure~\ref{fig:ssproser} shows the stellar surface density profiles of massive SDSS ($z<0.1$) galaxies in the sample presented by \citet{szomoru_etal12} rescaled using $r_n$ as above. Note that I plot not the actual measured profiles but the S\'ersic profiles with parameters derived from $M_*$ and $R_e$ values in Table 1 in that paper. The mean profile of late type galaxies from the top panel of Fig.~2 is shown for comparison. The figure shows that stellar distribution of early type galaxies also follows an approximately universal profile. The mean profile is very close to the de Vaucouleurs profile with $R_e=0.015\Rtwoh/1.34$, where factor of 1.34 converts 3D half-mass radius to the 2D $R_e$. Remarkably, the mean profiles of late type and early type galaxies are quite similar at $R\gtrsim r_n$ and are only significantly different within the half-mass radius. This implies that the same process shapes stellar distribution at large radii in both late and early type galaxies.   

\section{Discussion and conclusions}
\label{sec:conclusions}

Figures~\ref{fig:rhR200}-\ref{fig:ssproser} demonstrate that characteristic size of stellar and gas distributions in galaxies spanning more than eight orders of magnitude in stellar mass scales approximately linearly with the virial radius $\Rtwoh$ derived using abundance matching approach. Such relation is in remarkable agreement with expectations of the model of \citet{mo_etal98}: $\rhalf\propto\lambda\Rtwoh$. 

The agreement is not only qualitative, but also quantitative. The scatter in size at fixed $\Rtwoh$ in such model is mostly due to scatter in $\lambda$ -- the spin of parent halo. The dotted lines in Figure~\ref{fig:rhR200} show the scatter of $0.5$ dex, which approximately corresponds to $2\sigma_{\ln\lambda}\approx 1.1$, where $\sigma_{\ln\lambda}\approx 0.55$ is the rms width of the log-normal spin pdf \citep[e.g.,][]{vitvitska_etal02}. The figure shows that there is a good general agreement between the scatter due to the spin distribution and observed scatter of $\rhalf$ estimated for massive galaxies in the samples of \citet{szomoru_etal12} and \citet{bernardi_etal12}. The agreement is noteworthy, although more detailed comparison taking into account  measurement errors of $R_e$ and scatter of $\Rtwoh$ due to intrinsic scatter  in the $M_*-\Mtwoh$ relation and measurement errors of $M_*$ is needed.

\citet{mo_etal98} model predicts $\rhalf=1.678R_d=1.187(j_d/m_d)f_c^{-1/2}f_R\lambda\Rtwoh$, where $j_d$ and $m_d$ are fractions of baryon angular momentum and mass budget within halo in the central disk, $f_c$ is a function of halo concentration, and $f_R$ is a function that takes into account baryonic contraction of halo in response to halo formation  (see their \S~2.3). Assuming $j_d/m_d=1$, $m_d=0.05$, and typical spin $\bar{\lambda}=0.045$ and halo concentration of $c_{200}=10$ gives $\rhalf=0.032\Rtwoh$, i.e. normalization about a factor of two higher than inferred in this study. This, of course, could simply be due to $j_d/m_d<1$. However, as noted by \citet{mo_etal98}, the predicted relation is applicable at the epoch of disk formation. Indeed, after the galaxy size is set, $\Rtwoh$ may increase significantly simply due to pseudo-evolution of halo mass defined with respect to the evolving reference  density \citep{diemer_etal12}. If I assume that galaxy size was set at a characteristic epoch of $z\approx 2$ and that halo mass of most galaxies increases by a factor of 2.5 due primarily to pseudo-evolution between $z=2$ and $z=0$, as suggested by analysis of cosmological simulations \citep{diemer_etal12}, the halo radius that set the galaxy size is $\Rtwoh(z=2)\approx 0.37\Rtwoh(z=0)$. Using again $\bar{\lambda}=0.045$ and $c_{200}=4$ typical for $z=2$ halos, the prediction is $\rhalf\approx 0.016(j_d/m_d)\Rtwoh(z=0)$ in agreement with the relation derived for observed galaxies. This conjecture can, in principle, be tested via analysis similar to the one presented in this paper but done at $z=1-2$. 

These estimates demonstrate that empirically derived $\rhalf-\Rtwoh$ relation is in very good {\it quantitative} agreement with predictions of the \citet{mo_etal98} disk formation model. Remarkably, the prediction works not only for late type disks, but also for early type galaxies. This means that angular momentum plays a critical role in setting the sizes of galaxies of all morphological types. This fact reveals yet another remarkable regularity in properties of observed galaxies and provides a critical test for models of galaxy formation. 

The derived relation may also provide a useful way to estimate galaxy sizes in simulations when only halo information is available. Conversely, it can be used to derive halo extent and mass using the observed $\Reff$. As shown recently by \citet{szomoru_etal12}, half-light radius of galaxies is offset only by $\approx 25\%$ from half-mass radius $\rhalf$ regardless of galaxy stellar mass, morphology, and redshift. The relation derived in this study can thus be used to estimate $R_{200}$ of galaxy halos with $\approx 50\%$ accuracy and virial mass $M_{200}$ to within a factor of about four from the measurement of half-light radius alone without resorting to estimate of stellar mass. For these reasons it would be interesting to calibrate the relation and its scatter using larger, well-defined samples at a variety of redshifts. 

\acknowledgements
I am grateful to Mariangela Bernardi for useful discussions and communicating scatter results for her sample and to Nick Gnedin and Andrew Hearin for useful comments on the manuscript. 
This work was supported in part by the Kavli Institute for Cosmological Physics at the University of Chicago through grants NSF PHY-0551142 and PHY-1125897 and an endowment from the Kavli Foundation and its founder Fred Kavli. 

\bibliographystyle{apj}
\bibliography{ms}

\end{document}

%% file: def.tex
\newcommand{\Msun}{M_{\odot}}
\renewcommand{\vec}[1]{\bmath{#1}}
\newcommand{\be}{\begin{equation}}
\newcommand{\ee}{\end{equation}}
\newcommand{\ba}{\begin{eqnarray}}
\newcommand{\ea}{\end{eqnarray}}
\newcommand{\brr}{\begin{array}}
\newcommand{\err}{\end{array}}
\newcommand{\bc}{\begin{center}}
\newcommand{\ec}{\end{center}}
\newcommand{\hm}{\,h^{-1}{\rm Mpc}}
\newcommand{\hk}{\,h^{-1}{\rm kpc}}
\newcommand{\bx}{{\bf x}}
\newcommand{\msun}{\,h^{-1}M_\odot}
\newcommand{\hMpc}{\mbox{$h^{-1}{\rmn{Mpc}}~$}}
\newcommand{\rvir}{\mbox{$R_{\rmn{vir}}$}}
\newcommand{\mvir}{\mbox{$M_{\rmn{vir}}$}}
\newcommand{\Rdc}{\mbox{$R_{\Delta_c}$}}
\newcommand{\Rtwoh}{\mbox{$R_{200}$}}
\newcommand{\Mtwoh}{\mbox{$M_{200}$}}
\newcommand{\Dc}{\mbox{$\Delta_c$}}
\newcommand{\dc}{\mbox{$\delta_c$}}
\newcommand{\Dm}{\mbox{$\Delta_{\rm m}$}}
\newcommand{\Mdc}{\mbox{$M_{\Delta_c}$}}
\newcommand{\Mfhm}{\mbox{$M_{500}$}}
\newcommand{\Mfh}{M_{500}}
\newcommand{\Mdm}{\mbox{$M_{\Delta_{\rm m}}$}}
\newcommand{\Rd}{\mbox{$R_{\Delta}$}}
\newcommand{\rhalf}{\mbox{$r_{1/2}$}}
\newcommand{\Rhlf}{\mbox{$R_{1/2}$}}
\newcommand{\re}{\mbox{$r_{\rm e}$}}
\newcommand{\Reff}{\mbox{$R_{\rm e}$}}
\newcommand{\Md}{\mbox{$M_{\Delta}$}}
\newcommand{\Mgd}{\mbox{$M_{\rm g\Delta}$}}
\newcommand{\tmw}{\mbox{$T_{\rmn{mw}}$}}
\newcommand{\hMpcI}{\mbox{$h\,{\rmn{Mpc}}^{-1}$}}
\newcommand{\lb}{{\left<\right.}}
\newcommand{\rb}{{\left.\right>}}
\newcommand{\lum}{\,{\rm erg\,s^{-1}}}
\newcommand{\vel}{\,{\rm km\,s^{-1}}}
\newcommand{\hub}{\,{\rm km\,s^{-1}Mpc^{-1}}}
\newcommand{\lt}{$L_X$--$T$$~$}
\newcommand{\mincir}{\raise
  -2.truept\hbox{\rlap{\hbox{$\sim$}}\raise5.truept \hbox{$<$}\ }}
\newcommand{\magcir}{\raise
  -2.truept\hbox{\rlap{\hbox{$\sim$}}\raise5.truept \hbox{$>$}\ }}
\newcommand{\siml}{\raise
  -2.truept\hbox{\rlap{\hbox{$\sim$}}\raise5.truept \hbox{$<$}\ }}
\newcommand{\simg}{\raise
  -2.truept\hbox{\rlap{\hbox{$\sim$}}\raise5.truept \hbox{$>$}\ }}
\newcommand{\Mg}{M_{\rm g}}
\newcommand{\Mnl}{M_{\rm NL}}
\newcommand{\Mgas}{\mbox{$M_{\rm g}$}}
\newcommand{\fgas}{\mbox{$f_{\rm g}$}}
\newcommand{\Mstar}{\mbox{$M_{\ast}$}}
\newcommand{\Cg}{\mbox{$C_{\rm g}$}}
\newcommand{\Cs}{\mbox{$C_{\ast}$}}
\newcommand{\CT}{\mbox{$C_{\rm T}$}}
\newcommand{\ag}{\alpha_{\rm g}}
\newcommand{\as}{\alpha_{\ast}}
\newcommand{\aT}{\alpha_{\rm T}}
\newcommand{\Cgo}{\mbox{$C_{\rm g0}$}}
\newcommand{\CTo}{\mbox{$C_{\rm T0}$}}
\newcommand{\Cso}{\mbox{$C_{\ast0}$}}
\newcommand{\rhog}{\rho_{\rm g}}
\newcommand{\rhogs}{\rho_{\rm g\ast}}
\newcommand{\trhog}{\tilde{\rho}_{\rm g}}
\newcommand{\Ts}{T_{\ast}}
\newcommand{\Tx}{T_{\rm X}}
\newcommand{\tT}{\tilde{T}}
\newcommand{\tphi}{\tilde{\phi}}
\newcommand{\Lx}{L_{\rm X}}
\newcommand{\Lsx}{L_{\rm Xs}}
\newcommand{\Lbol}{L_{\rm bol}}
\newcommand{\LCh}{L_{\rm Ch}}
\newcommand{\Omb}{\Omega_{\rm b}}
\newcommand{\Omm}{\Omega_{\rm m}}
\newcommand{\Oml}{\Omega_{\rm \Lambda}}
\newcommand{\Omx}{\Omega_{\rm X}}
\newcommand{\Omde}{\Omega_{\rm DE}}
\newcommand{\rhom}{\rho_{\rm m}}
\newcommand{\rhoc}{\rho_{\rm cr}}
\newcommand{\rhoco}{\rho_{\rm cr0}}

%% file: macros.tex
%
%
%

\newcommand{\etal}{{et al.~}}

\newcommand{\kmsmpc}{\>{\rm km}\,{\rm s}^{-1}\,{\rm Mpc}^{-1}}
\newcommand{\kms}{\>{\rm km}\,{\rm s}^{-1}}
\newcommand{\pc}{\>{\rm pc}}
\newcommand{\cm}{\>{\rm cm}}
\newcommand{\Mpc}{\>{\rm Mpc}}
\newcommand{\kpc}{\>{\rm kpc}}
\newcommand{\rMsun}{\>{\rm M_{\odot}}}
\newcommand{\rLsun}{\>{\rm L_{\odot}}}
\newcommand{\MLsun}{\>({\rm M}/{\rm L})_{\odot}}
\newcommand{\Mbh}{M_{\bullet}}
\newcommand{\Vrot}{V_{\rm rot}}
\newcommand{\mtol}{\>{\rm (M/L)_{\odot}}}
\newcommand{\erg}{\>{\rm erg}}
\newcommand{\kpch}{\>{h^{-1}{\rm kpc}}}
\newcommand{\mpch}{\>h^{-1}{\rm {Mpc}}}
\newcommand{\yr}{\>{\rm yr}}
\newcommand{\yrs}{\>{\rm yrs}}
\newcommand{\Msunh}{\>h^{-1}\rm M_\odot}
\newcommand{\Lsunh}{\>h^{-2}\rm L_\odot}
\newcommand{\calN}{{\cal N}}
\newcommand{\wcalN}{\tilde{{\cal N}}}
\newcommand{\walpha}{\tilde{\alpha}}
\newcommand{\wLstar}{\tilde{L}^{*}}
\newcommand{\hxi}{\hat{\xi}}
\newcommand{\lamA}{${\Lambda}30/90 \, $}
\newcommand{\lamC}{${\Lambda}25/75 \, $}
\newcommand{\lamD}{${\Lambda}20/65 \, $}
\newcommand{\lamB}{${\Lambda}30/65 \, $}
\newcommand{\beq}{\begin{equation}}
\newcommand{\eeq}{\end{equation}}
\newcommand{\vcir}{V_{\rm c}}
\newcommand{\vh}{V_{\rm c}}
\newcommand{\Obaryon}{{\Omega_{\rm B,0}}}
\newcommand{\Kdegree}{\>{\rm K}}
\newcommand{\keV}{\>{\rm keV}}
\newcommand{\vhalo}{V_{\rm c}}
\newcommand{\Tvir}{T_{\rm vir}}
\newcommand{\rmd}{{\rm d}}
\newcommand{\vesc}{V_{\rm esc}}
\newcommand{\Lya}{{\rm Ly}\alpha}
\newcommand{\msunh}{\>h^{-1}\rm M_\odot}
\newcommand{\Lsunhh}{\,h^{-2}\rm L_\odot}

\newcommand{\lcen}{L_{\rm c}}
\newcommand{\avg}[1]{\langle #1 \rangle}
\newcommand{\avglogm}{\avg{\log M}(\lcen)}
\newcommand{\avgloglc}{\avg{\log \lcen}(M)}
\newcommand{\avglogsqlc}{\avg{(\log \lcen)^2}(M)}
\newcommand{\siglogm}{\sigma_{\log M}(\lcen)}
\newcommand{\sigint}{\sigma_{\rm int}}
\newcommand{\ploglcm}{P(\log \lcen|M)}
\newcommand{\plogmlc}{P(\log M|\lcen)}
\newcommand{\drm}{{\rm d}}
\newcommand{\pdv}{{P (\Delta V)}}
\newcommand{\dv}{{\Delta V}}
\newcommand{\sigcen}{\sigma_{\log L}}
\newcommand{\sigsw}{\sigma_{\rm sw}}
\newcommand{\sighw}{\sigma_{\rm hw}}
\newcommand{\sigsat}{\sigma_{\rm sat}}
\newcommand{\sigsatsq}{\sigma_{\rm sat}^2}
\newcommand{\avgsigsatsq}{\avg{\sigsat^2}}
\newcommand{\avnsat}{\avg{N_{\rm sat}}}
\newcommand{\avnsatm}{{\avnsat}_M}
\newcommand{\philcm}{\Phi_{\rm c}(L|M)}
\newcommand{\philsm}{\Phi_{\rm s}(L|M)}
\newcommand{\lc}{L_{\rm c}}
\newcommand{\plcm}{P(\lc|M)}
\newcommand{\pmlc}{P(M|\lc)}
\newcommand{\sigav}{\sigma_{\rm av}}
\newcommand{\vecb}[1]{{\bf #1}}
\newcommand{\bs}[1]{{\boldsymbol #1}}
\newcommand{\e}[1]{\vecb{\hat e}_{#1}}
\newcommand{\sqterm}[2]{#1^T#2#1}
\newcommand{\xyterm}[2]{#1^T#2+#2^T#1}
\newcommand{\erf}{{\rm erf}}
\newcommand{\mgas}{M_{\rm gas}}
\newcommand{\tx}{T_{\rm x}}
\newcommand{\mwl}{M_{\rm wl}}
\newcommand{\mfive}{M_{500}}
\newcommand{\eM}{\epsilon_{\rm M}}
\newcommand{\elss}{\epsilon_{\rm E}}
\newcommand{\eT}{\epsilon_{\rm T}}
\newcommand{\eMg}{\epsilon_{\rm g}}


\def\gtsima{$\; \buildrel > \over \sim \;$}
\def\ltsima{$\; \buildrel < \over \sim \;$}
\def\prosima{$\; \buildrel \propto \over \sim \;$}
\def\gsim{\lower.7ex\hbox{\gtsima}}
\def\lsim{\lower.7ex\hbox{\ltsima}}
\def\simgt{\lower.7ex\hbox{\gtsima}}
\def\simlt{\lower.7ex\hbox{\ltsima}}
\def\simpr{\lower.7ex\hbox{\prosima}}
\def\la{\lsim}
\def\ga{\gsim}
\def\lta{\la}
\def\gta{\ga}


\newcommand{\XXX}[2]{{\sf #1}}
\newcommand{\QQQ}[1]{{\sc $<$#1$>$}}






\newdimen\hssize
\hssize=8.4truecm
\newdimen\hdsize
\hdsize=17.7truecm


\def\fn#1{$^{\ref{#1}}$}
\def\fit#1{\footnotesize \it #1 }
        

\def\red{\color{red}}
\def\blue{\color{blue}}
\def\rhohalo{\rho_{\rm halo}}
\def\mmean{M_{200\bar{\rho}}}
\def\rmean{R_{200\bar{\rho}}}
\def\mcrit{M_{500\rho_c}}
\def\rcrit{R_{500\rho_c}}
\def\mvir{M_{\rm vir}}
\def\rvir{R_{\rm vir}}


\def\rma{{\rm a}}
\def\rmb{{\rm b}}
\def\rmc{{\rm c}}
\def\rmd{{\rm d}}
\def\rme{{\rm e}}
\def\rmf{{\rm f}}
\def\rmg{{\rm g}}
\def\rmh{{\rm h}}
\def\rmi{{\rm i}}
\def\rmj{{\rm j}}
\def\rmk{{\rm k}}
\def\rmK{{\rm K}}
\def\rml{{\rm l}}
\def\rmm{{\rm m}}
\def\rmn{{\rm n}}
\def\rmo{{\rm o}}
\def\rmp{{\rm p}}
\def\rmq{{\rm q}}
\def\rmr{{\rm r}}
\def\rms{{\rm s}}
\def\rmt{{\rm t}}
\def\rmu{{\rm u}}
\def\rmv{{\rm v}}
\def\rmw{{\rm w}}
\def\rmx{{\rm x}}
\def\rmy{{\rm y}}
\def\rmz{{\rm z}}

\def\rmA{{\rm A}}
\def\rmB{{\rm B}}
\def\rmC{{\rm C}}
\def\rmD{{\rm D}}
\def\rmE{{\rm E}}
\def\rmF{{\rm F}}
\def\rmG{{\rm G}}
\def\rmH{{\rm H}}
\def\rmI{{\rm I}}
\def\rmJ{{\rm J}}
\def\rmK{{\rm K}}
\def\rmL{{\rm L}}
\def\rmM{{\rm M}}
\def\rmN{{\rm N}}
\def\rmO{{\rm O}}
\def\rmP{{\rm P}}
\def\rmQ{{\rm Q}}
\def\rmR{{\rm R}}
\def\rmS{{\rm S}}
\def\rmT{{\rm T}}
\def\rmU{{\rm U}}
\def\rmV{{\rm V}}
\def\rmW{{\rm W}}
\def\rmX{{\rm X}}
\def\rmY{{\rm Y}}
\def\rmZ{{\rm Z}}

\def\calA{{\cal A}}
\def\calB{{\cal B}}
\def\calC{{\cal C}}
\def\calD{{\cal D}}
\def\calE{{\cal E}}
\def\calF{{\cal F}}
\def\calG{{\cal G}}
\def\calH{{\cal H}}
\def\calI{{\cal I}}
\def\calJ{{\cal J}}
\def\calK{{\cal K}}
\def\calL{{\cal L}}
\def\calM{{\cal M}}
\def\calN{{\cal N}}
\def\calO{{\cal O}}
\def\calP{{\cal P}}
\def\calQ{{\cal Q}}
\def\calR{{\cal R}}
\def\calS{{\cal S}}
\def\calT{{\cal T}}
\def\calU{{\cal U}}
\def\calV{{\cal V}}
\def\calW{{\cal W}}
\def\calX{{\cal X}}
\def\calY{{\cal Y}}
\def\calZ{{\cal Z}}

\def\ba{{\bf a}}
\def\bb{{\bf b}}
\def\bc{{\bf c}}
\def\bd{{\bf d}}
\def\be{{\bf e}}
\def\bff{{\bf f}}
\def\bg{{\bf g}}
\def\bh{{\bf h}}
\def\bi{{\bf i}}
\def\bj{{\bf j}}
\def\bk{{\bf k}}
\def\bl{{\bf l}}
\def\bm{{\bf m}}
\def\bn{{\bf n}}
\def\bo{{\bf o}}
\def\bp{{\bf p}}
\def\bq{{\bf q}}
\def\br{{\bf r}}
\def\bs{{\bf s}}
\def\bt{{\bf t}}
\def\bu{{\bf u}}
\def\bv{{\bf v}}
\def\bw{{\bf w}}
\def\bx{{\bf x}}
\def\by{{\bf y}}
\def\bz{{\bf z}}

\def\bA{{\bf A}}
\def\bB{{\bf B}}
\def\bC{{\bf C}}
\def\bD{{\bf D}}
\def\bE{{\bf E}}
\def\bF{{\bf F}}
\def\bG{{\bf G}}
\def\bH{{\bf H}}
\def\bI{{\bf I}}
\def\bJ{{\bf J}}
\def\bK{{\bf K}}
\def\bL{{\bf L}}
\def\bM{{\bf M}}
\def\bN{{\bf N}}
\def\bO{{\bf O}}
\def\bP{{\bf P}}
\def\bQ{{\bf Q}}
\def\bR{{\bf R}}
\def\bS{{\bf S}}
\def\bT{{\bf T}}
\def\bU{{\bf U}}
\def\bV{{\bf V}}
\def\bW{{\bf W}}
\def\bX{{\bf X}}
\def\bY{{\bf Y}}
\def\bZ{{\bf Z}}